\documentclass[11pt]{amsart}

\usepackage{amscd,amssymb,amsmath,latexsym,enumerate}
\usepackage[utf8]{inputenc}
\usepackage{amscd,amssymb,amsmath,amsthm}
\usepackage[mathscr]{euscript}
\usepackage{mathrsfs}
\usepackage{hyperref}
\usepackage{graphicx}
\usepackage{color}

\usepackage{lscape} 		%to rotate pages
\usepackage{epstopdf}
\usepackage{enumitem}
\usepackage[dvipsnames]{xcolor}

\usepackage{changepage}

\usepackage{array}
\usepackage{tabularx}
\usepackage{longtable}
\usepackage{ltxtable}

\usepackage{color}

\usepackage{tikz}
\usetikzlibrary{decorations,arrows}
\usetikzlibrary{decorations.pathmorphing}
\usepgflibrary{decorations.pathreplacing}
\usepackage{pgf}
\usepackage{pgfkeys}
\usepackage{pgffor}

\usepackage{pgfpages}
\usepackage{scalefnt}
\usetikzlibrary{shapes}

\allowdisplaybreaks

\DeclareMathAlphabet{\mathpzc}{OT1}{pzc}{m}{it}

\newtheorem{definition}{Definition}
\newtheorem{theorem}[definition]{Theorem}
\newtheorem{proposition}[definition]{Proposition}

\newtheorem*{remark}{Remark}

\theoremstyle{remark}

\newcommand{\N}{{\mathbb N}}
\newcommand{\Q}{{\mathbb Q}}
\newcommand{\R}{{\mathbb R}}
\newcommand{\Z}{{\mathbb Z}}

\allowdisplaybreaks

\usepackage{xcolor}
\newcommand{\Hmm}[1]{\leavevmode{\marginpar{\tiny%
$\hbox to 0mm{\hspace*{-0.5mm}$\leftarrow$\hss}%
\vcenter{\vrule depth 0.1mm height 0.1mm width \the\marginparwidth}%
\hbox to 0mm{\hss$\rightarrow$\hspace*{-0.5mm}}$\\\relax\raggedright #1}}}

\begin{document}
\title{MFO Report: The dry ten Martini problem for Sturmian dynamical systems}
\author{Ram Band, Siegfried Beckus, Raphael Loewy}

\address{Department of Mathematics\\
Technion - Israel Institute of Technology\\
Haifa, Israel}
\email{ramband@technion.ac.il}

\address{Institute of Mathematics\\
University of Potsdam\\
Potsdam, Germany}
\email{beckus@uni-potsdam.de}

\address{Department of Mathematics\\
Technion - Israel Institute of Technology\\
Haifa, Israel}
\email{loewy@technion.ac.il}

\begin{abstract}
This extended Oberwolfach report (to appear in \cite{MFO_Bec23}) announces the full solution to the Dry Ten Martini Problem for Sturmian Hamiltonians \cite{BBL23}. Specifically, we show that all spectral gaps of Sturmian Hamiltonians (as predicted by the gap labeling theorem) are open for all nonzero couplings and all irrational rotations. We present here the proof strategy.
\end{abstract}

%%%%%%%%%%%%%%%%%%%%%%%%%%%%%%%%%%%%%%%%%%%%%%%%%%%%%%%%%%%%%%%%%%%%
\maketitle

Are all possible spectral gaps, predicted by the gap labeling theorem, open for a given Schr\"odinger operator?  This is the so called {\em Dry Ten Martini problem (Dry TMP)} motivated by the {\em Ten Martini Problem (TMP)}. The name TMP was coined by Simon \cite{Sim82} after Kac offered in 1981 ten Martinis to anyone who solves it. Originally, the TMP was proposed for the Almost Mathieu operator conjecturing Cantor spectrum for all couplings and all irrational frequencies. The TMP for the Almost Mathieu operator was solved by Avila and Jitomirskaya \cite{AviJit05}. Recently, Avila, You and Zhou \cite{AviYouZho23} also solved the Dry Ten Martini Problem in the non-critical case for the Almost-Mathieu operator.

\medskip

\begin{figure}[bt]
\includegraphics[scale=0.135]{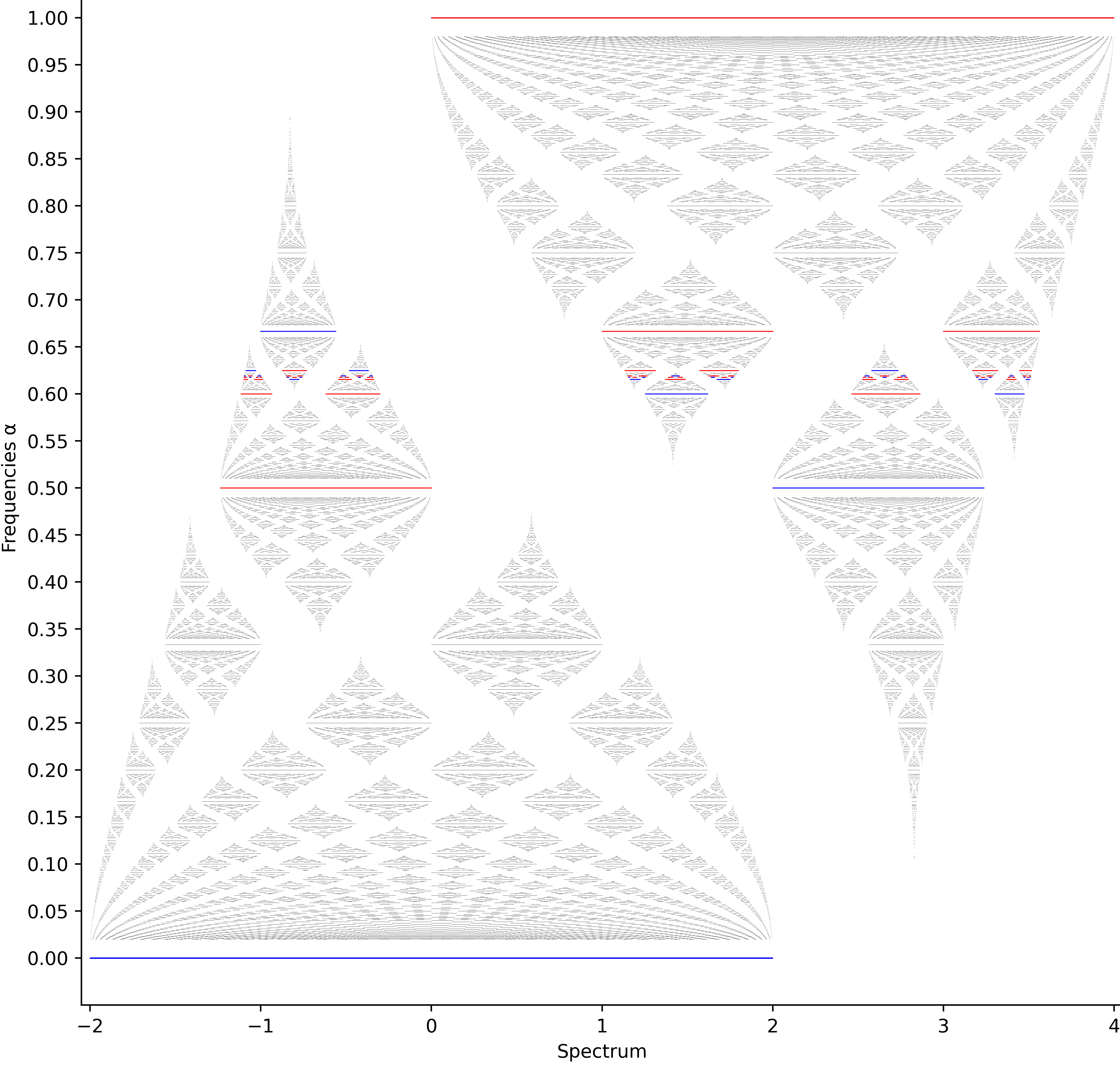}
\caption{Kohmoto butterfly for $V=2$: The vertical axis represents the frequency values $\alpha$. For each $\alpha$ the spectrum of $H_{\alpha,V}$ is plotted horizontally. A few spectra $H_{\frac{p}{q},V}$ are colored for $\frac{p}{q}\in\big\{ \frac{0}{1}, \frac{1}{1}, \frac{1}{2}, \frac{2}{3}, \frac{3}{5}, \frac{5}{8} \big\}$ representing the first periodic approximations of the Fibonacci Hamiltonian. According to Definition~\ref{Def-TypeAB}, $A$-type bands ($B$-type bands) are colored blue (red).
}
\label{Fig-KohmotoButterfly}
\end{figure}

Sturmian dynamical systems are closely related dynamical systems while reflecting the aperiodicity coming from a solid and not from a magnetic field. These systems are used to define a one-dimensional Schr\"odinger operator $H_{\alpha,V}:\ell^2(\Z)\to\ell^2(\Z)$,
$$
(H_{\alpha,V}\psi)(n) = \psi(n-1) + \psi(n+1) + V\chi_{[1-\alpha,1)}(\{n\alpha\}) \psi(n), \qquad \psi\in\ell^2(\Z),\, n\in\Z,
$$
where $\{n\alpha\} := n\alpha-\lfloor n\alpha\rfloor$ is the fractional part of $n\alpha$. The potential above is characterized in terms of two parameters: a frequency parameter $\alpha\in[0,1]$ and the potential strength, also known as the coupling constant $V\in\R$. Since $H_{\alpha,V}$ is a linear, self-adjoint and bounded operator, its spectrum $\sigma(H_{\alpha,V})$ is a compact subset of $\R$. The family of Schr\"odinger operators $H_{\alpha,V}$ for $\alpha\in[0,1]\setminus\Q$ and $V\neq 0$ is called {\em Sturmian Hamiltonians} or {\em Kohmoto model}. The so-called {\em Kohmoto butterfly} is the corresponding plot of the spectra of $H_{\alpha,V}$ as it varies with $\alpha$, see Figure~\ref{Fig-KohmotoButterfly}.

The integrated density of states (IDS) $N_{\alpha,V}:\R\to [0,1]$ is defined by
$$
N_{\alpha,V}(E) := \lim_{n\to\infty} \frac{\sharp \big\{ \lambda \in \sigma\big(H_{\alpha,V}|_{[1,n]}\big) \,\big|\, \lambda\leq E\big\}}{n}, \qquad E\in\R,
$$
where $H_{\alpha,V}|_{[1,n]}$ is the $n\times n$ matrix obtained by restricting $H_{\alpha,V}$ to $\ell^2(1,\ldots,n)$, see e.g. \cite{BeBoGh91-Gap,DamFil22} for more details. 
In particular, the IDS for Sturmian Hamiltonians satisfies the following properties:
\begin{enumerate}
\item[(IDS1)] $N_{\alpha,V}:\R\to[0,1]$ is monotone increasing and continuous, 
\item[(IDS2)] $E\in \R\setminus \sigma(H_{\alpha,V})$ if and only if there exists an $\varepsilon>0$ such that the restriction $N_{\alpha,V}|_{(E-\varepsilon,E+\varepsilon)}$ is constant. 
\end{enumerate}
A connected component $g:=(a,b)$ in $\R\setminus \sigma(H_{\alpha,V})$ is called a {\em spectral gap}, i.e. $a,b\in\sigma(H_{\alpha,V})$ and $(a,b)\cap \sigma(H_{\alpha,V})=\emptyset$. By (IDS1) and (IDS2), we have $N_{\alpha,V}(E)=N_{\alpha,V}(E')$ for all $E,E'\in[a,b]$. The value $N_{\alpha,V}(E)$ for some (any) $E\in g$ is called the {\em gap label} of $g$. The gap labeling theorem \cite{BeBoGh91-Gap} prescribes all the possible labels that the spectral gaps of $H_{\alpha,V}$ may have. Specifically, for Sturmian Hamiltonians, we have
$$
%\{\tau(g) \,|\, g \text{ spectral gap of } H_{\alpha,V}\} = 
	\{ N_{\alpha,V}(E) \,|\, E\in \R\setminus \sigma(H_{\alpha,V})\}
	\subseteq \big\{ \{n\alpha\} \,|\, n\in\Z\big\} \cup \{1\},
$$
where $\{n\alpha\}$ is the fractional part of $n\alpha$ as above. One may ask whether the latter inclusion is an equality - this is the Dry Ten Martini Problem for Sturmian dynamical systems. A complete solution of this problem appears in \cite{BBL23} (see also \cite{MWL23} for the first announcement).

\begin{theorem}[Sturmian Dry TMP] 
\label{thm-DTMP}
For all $\alpha\in[0,1]\setminus\Q$ and all $V\neq 0$, all possible spectral gaps of $H_{\alpha,V}$ are open, i.e.
$$
\{ N_{\alpha,V}(E) \,|\, E\in \R\setminus \sigma(H_{\alpha,V})\}
	= \big\{ \{n\alpha\} \,|\, n\in\Z\big\} \cup \{1\}.
$$
\end{theorem}

Let us shortly discuss the progress that has been done so far towards a resolution of the Sturmian Dry TMP. For large couplings, $V>4$, the previous theorem was proven by Raymond \cite{Raym} (see also \cite{BBBT23} for a review).
For the Fibonacci Hamiltonian ($\alpha$ equals to the golden mean $\varphi:=\frac{\sqrt{5}-1}{2}$), Damanik and Gorodetski proved that all gaps are open for sufficiently small values of the coupling constant $V$, see \cite[Theorem~1.5]{DaGo11}.
This (small coupling result) was extended by Mei in \cite[Theorem~1.5]{Mei14} to all $\alpha\in[0,1]\setminus \Q$ with eventually periodic continued fraction expansion. 
Later, Damanik, Gorodetski and Yessen provided a complete solution for the Fibonacci Hamiltonian in \cite[Theorem~1.3]{DaGoYe16}. One might extend this result to all irrational $\alpha$'s with eventually periodic continued fraction expansion, confer \cite[Section~7]{DaGoYe16}. In the latter work it was also conjectured that the Sturmian Dry TMP is true for all irrational $\alpha\in[0,1]$ and couplings $V\neq 0$.

\medskip

The rest of the note presents the main strategy of the proof of Theorem~\ref{thm-DTMP} in five steps.

\subsection*{Step I: Spectra of periodic approximations}

Let $\alpha\in[0,1]\setminus\Q$. Then the continued fraction expansion of $\alpha$ is given by the unique sequence $a_k\in\N$ for $k\geq 0$ satisfying
$$
\alpha = a_0 + \frac{1}{a_1 + \frac{1}{a_2 + \frac{1}{\ddots}}} =:[a_0,a_1,a_2,\ldots].
$$
For $k\geq 0$, we also consider the Diophantine approximations $\alpha_k := [0,a_1,\ldots,a_k]= \frac{p_k}{q_k}$ (where $p_k,q_k\in\N$ are coprime). We refer to these as finite continued fraction expansions.

\medskip

Since $\alpha_k\in\Q$, the operator $H_{\alpha_k,V}$ is periodic. Thus, the spectrum $\sigma(H_{\alpha_k,V})$ is a finite union of exactly $q_k$ intervals by the Floquet-Bloch theory. These intervals of $\sigma(H_{\alpha_k,V})$ are called {\em spectral bands}. Moreover, we have the following result \cite{BIST89,BIT91,Raym}.

\begin{proposition}[Spectral approximations]
\label{prop-SpectralApprox}
Let $V> 0$, $\alpha\in[0,1]\setminus\Q$ and $\alpha_k= [0,a_1,\ldots,a_k]$ for $k\in\N$. Define the compact sets
$$
\Sigma_{k,V} := \sigma(H_{\alpha_k,V}) \cup \sigma(H_{\alpha_{k+1},V})\subseteq \R,\qquad k\in\N.
$$
Then $\Sigma_{k+1,V}\subseteq \Sigma_{k,V}$ and
$$
\lim_{k\to\infty} \Sigma_{k,V} = \bigcap_{k\in\N} \Sigma_{k,V} = \sigma(H_{\alpha,V})
$$
where the limit is taken with respect to the Hausdorff metric on the compact subsets of $\mathbb{R}$.
\end{proposition}

We refer the reader to Figure~\ref{Fig-KohmotoButterfly}, which demonstrates the first few periodic approximations of the Fibonacci Hamiltonian ($\alpha=\varphi=[0,1,1,1,1,\ldots]$).

\begin{remark}
A more general result is proven in \cite{Raym}. Here, we provide a simplified version, which is sufficient to present our proof strategy.
\end{remark}

\medskip

\subsection*{Step II: Combinatorial structure of the spectral approximations}

The monotone convergence in Proposition~\ref{prop-SpectralApprox} allows to classify the spectral bands into two types.

\begin{definition}
\label{Def-TypeAB}
A spectral band $I(V)$ in $\sigma(H_{\alpha_k,V})$ is called
\begin{itemize}
\item of {\em type A} if it is strictly contained in a spectral band of $\sigma(H_{\alpha_{k-1},V})$;
\item of {\em type B} if it is not of type A and it is strictly contained in a spectral band of $\sigma(H_{\alpha_{k-2},V})$;
\end{itemize}
\end{definition}

\begin{remark}
We note that the analogous notions of types $A$ and $B$ appearing in \cite{BBBT23,BBL23} are different than those in Definition~\ref{Def-TypeAB}, even though both are equivalent. The different definition there is essential to prove the next theorem.
\end{remark}

\begin{figure}[htb]
\begin{tikzpicture}[>=stealth,scale=0.82, every node/.style={scale=0.82}]

%%%%%%%%%%%%%%%%%%%%%%%%%%%%%%%%%%%%%%%%%%%%%%%%

\node at (-4.5,3.7) {(a)};
\node at (4.2,3.7) {(b)};

\node at (-4,3) {\footnotesize $\alpha_1$};
%\filldraw (-2,3) circle (1pt);
\node[red] at (2,3) {\footnotesize $B$};
\draw[thick] (1.3,2.8) -- (2.7,2.8);
\draw (1.3,2.85)--(1.3,2.75);
\draw (2.7,2.85)--(2.7,2.75);

%%%%%
%\draw[dotted] (-2,1.7)--(-2,3);
\draw[->] (2,1.3)--(2,2.8);
%%%%%

\node at (-4,1.5) {\footnotesize $\alpha_0$};
\node[blue] at (-2,1.5) {\footnotesize $A$};
\draw[thick] (-3,1.3) -- (-1,1.3);
\node at (-3.05,1.1) {\tiny $-2$};
\draw (-3,1.35)--(-3,1.25);
\node at (-1,1.1) {\tiny $2$};
\draw (-1,1.35)--(-1,1.25);

%\filldraw[thick] (3,1.5) circle (1pt);
%\draw[thick] (2,1.3) -- (4,1.3);
%\node at (3.95,1.1) {\tiny $2+V$};
%\draw (4,1.35)--(4,1.25);
%\node at (2,1.1) {\tiny $-2+V$};
%\draw (2,1.35)--(2,1.25);

%%%%%
\draw[->] (0,0) -- (-2,1.25);
\draw (0,0) -- (2,1.3);
%%%%%

\node at (-4,0) {\footnotesize root};
\filldraw (0,0) circle (1pt);
\draw[thick, dotted] (-3.5,0) -- (3.5,0);
\draw[thick] (-3,0) -- (3,0);
\node at (1,-0.3) {\footnotesize $\mathbb{R}$};

\draw[ultra thick,gray, opacity=0.5] (3.7,3.9)--(3.7,-0.5);

%%%%%%%%%%%%%%%%%%%%%%%%%%%%%%%%%%%%%%%%%%
%%%%%%%%%%%%%%%%%%%%%%%%%%%%%%%%%%%%%%%%%%
%%forward property
%%%%%%%%%%%%%%%%%%%%%%%%%%%%%%%%%%%%%%%%%%
%%%%%%%%%%%%%%%%%%%%%%%%%%%%%%%%%%%%%%%%%%

\begin{scope}[shift={(9,0)}]
\node at (-4,3) {\footnotesize $\alpha_{k+2}$};
%\filldraw (-2,3) circle (1pt);

%1 B-band
\node[red] at (-3,3) {\footnotesize $B$};
\node[Gray] at (-2.6,2.55) {\footnotesize $K^{(1)}$};
\draw[thick] (-3.3,2.8) -- (-2.7,2.8);
\draw (-3.3,2.85)--(-3.3,2.75);
\draw (-2.7,2.85)--(-2.7,2.75);

%2 B-band
\node[red] at (-1.2,3) {\footnotesize $B$};
\node[Gray] at (-0.7,2.55) {\footnotesize $K^{(2)}$};
\draw[thick] (-1.8,2.8) -- (-0.7,2.8);
\draw (-1.8,2.85)--(-1.8,2.75);
\draw (-0.7,2.85)--(-0.7,2.75);

%3 B-band
\node[red] at (1,3) {\footnotesize $B$};
\node[Gray] at (1.7,2.55) {\footnotesize $K^{(3)}$};
\draw[thick] (0.3,2.8) -- (1.7,2.8);
\draw (0.3,2.85)--(0.3,2.75);
\draw (1.7,2.85)--(1.7,2.75);

%%%%%
\draw[dotted] (2,2.8)--(2.5,2.8);
%%%%%

%p B-band
\node[red] at (3,3) {\footnotesize $B$};
\node[Gray] at (3.45,2.55) {\footnotesize $K^{(p)}$};
\draw[thick] (2.65,2.8) -- (3.35,2.8);
\draw (2.65,2.85)--(2.65,2.75);
\draw (3.35,2.85)--(3.35,2.75);

%p+1 B-band
\node[red] at (4.6,3) {\footnotesize $B$};
\node[Gray] at (5.4,2.55) {\footnotesize $K^{(p+1)}$};
\draw[thick] (3.9,2.8) -- (5.3,2.8);
\draw (3.9,2.85)--(3.9,2.75);
\draw (5.3,2.85)--(5.3,2.75);

%%%%%%%%%%%%%%%%%%%%%%%%%%%%%%%%%%%%%%%%%%%%%%%%%%%%%%%%%%%%%%%%%%%%%%%%%%%%%%%%%%%%%

\node at (-4,1.5) {\footnotesize $\alpha_{k+1}$};
\node[blue] at (-2,1.5) {\footnotesize $A$};
\node[Gray] at (-1.5,1.5) {\footnotesize $J^{(1)}$};
%1 A-band
\draw[thick] (-2.4,1.3) -- (-1.6,1.3);
\draw (-2.4,1.35)--(-2.4,1.25);
\draw (-1.6,1.35)--(-1.6,1.25);
%2 A-band
\node[blue] at (-0.15,1.5) {\footnotesize $A$};
\node[Gray] at (0.5,1.5) {\footnotesize $J^{(2)}$};
\draw[thick] (-0.8,1.3) -- (0.5,1.3);
\draw (-0.8,1.35)--(-0.8,1.25);
\draw (0.5,1.35)--(0.5,1.25);
%%%%%
\draw[dotted] (1.5,1.3)--(2.6,1.3);
%%%%%
%p A-band
\node[blue] at (3.7,1.5) {\footnotesize $A$};
\node[Gray] at (4.25,1.5) {\footnotesize $J^{(p)}$};
\draw[thick] (3.4,1.3) -- (4,1.3);
\draw (3.4,1.35)--(3.4,1.25);
\draw (4,1.35)--(4,1.25);

%%%%%%%%%%%%%%%%%%%%%%%%%%%%%%%%%%%%%%%%%%%%%%%%%%%%%%%%%%%%%%%%%%%%%%%%%%%%%%%%%%%%%

%%%%%
%1 B-band
\draw (0.75,0) -- (-3,1.3);
\draw[->] (-3,1.3)--(-3,2.75);
%1 A-band
\draw[->] (0.75,0) -- (-2,1.25);

%2 B-band
\draw (0.75,0) -- (-1.2,1.3);
\draw[->] (-1.2,1.3)--(-1.2,2.75);
%2 A-band
\draw[->] (0.75,0) -- (-0.15,1.25);

%3 B-band
\draw (0.75,0) -- (1,1.3);
\draw[->] (1,1.3)--(1,2.75);
%%%%%

%%%%%
\draw[dotted] (1.1,0.65)--(1.6,0.65);
%%%%%

%p B-band
\draw (0.75,0) -- (3,1.3);
\draw[->] (3,1.3)--(3,2.75);
%p A-band
\draw[->] (0.75,0) -- (3.7,1.25);
%p+1 B-band
\draw (0.75,0) -- (4.6,1.3);
\draw[->] (4.6,1.3)--(4.6,2.75);

\node at (-4,0) {\footnotesize $\alpha_k$};
\filldraw (0.75,0) circle (1pt);
\draw[thick] (-3.5,0) -- (5.2,0);
\draw (-3.5,-0.05)--(-3.5,0.05);
\draw (5.2,-0.05)--(5.2,0.05);
\node[blue] at (0.75,-0.3) {\footnotesize $A \ \textcolor{black}{/} \ \textcolor{red}{B}$};
\node[gray] at (3.75,-0.3) {\footnotesize $I(V)$};
\end{scope}
\end{tikzpicture}
\caption{(a) The root of the tree $T$ and two adjacent vertices. (b) The interval $I(V)$ of $\sigma(H_{\alpha_k,V})$ and the spectral bands of $\sigma(H_{\alpha_{k+1},V})$ and $\sigma(H_{\alpha_{k+2},V})$ that it includes.}
\label{Fig-TreeData}
\end{figure}
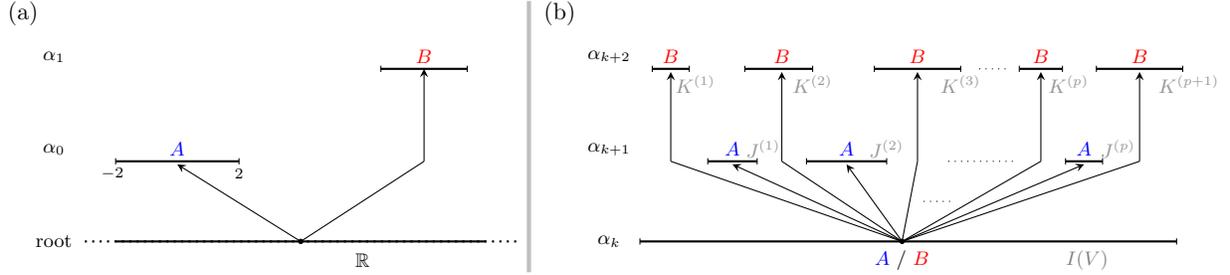

To state the next result, we define an order relation on spectral bands. Specifically, $[a,b]\prec [c,d]$ holds if $a<c$ and $b<d$. With this at hand, we have the following (see Figure~\ref{Fig-TreeData}~(b)).

\begin{theorem}
\label{thm-TypeA+B}
Let $V> 0$, $\alpha\in[0,1]\setminus\Q$ and $\alpha_k= [0,a_1,\ldots,a_k]$ for $k\in\N$. Then the following assertions hold. 
\begin{enumerate}
\item[(a)] Each spectral band $I(V)$ in $\sigma(H_{\alpha_k,V})$ is either of type A or of type B. 
\item[(b)] Let $I(V)$ be a spectral band in $\sigma(H_{\alpha_k,V})$ and
$$
p:= 
	\begin{cases}
		a_{k+1} - 1, \qquad &\text{if $I(V)$ is of type $A$},\\
		a_{k+1}, \qquad &\text{if $I(V)$ is of type $B$}.
	\end{cases}
$$
Then there exist
	\begin{itemize}
	\item $p$ spectral bands $J^{(1)},\ldots, J^{(p)}$ of type A in $\sigma(H_{\alpha_{k+1},V})$, and
	\item $p+1$ spectral bands $K^{(1)},\ldots, K^{(p+1)}$ of type B in $\sigma(H_{\alpha_{k+2},V})$ 
	\end{itemize}	 
that are all strictly contained in $I(V)$ and they interlace 
$$
K^{(j)}\prec J^{(j)}\prec K^{(j+1)}, \qquad 1\leq j \leq p.
$$
\end{enumerate}
\end{theorem}

\begin{remark}
We point out that in the interlacing property spectral bands may overlap (as in Figure~\ref{Fig-TreeData}~(b)) for small couplings $V$, which is not the case in \cite{Raym}. 
These overlaps are the source of the difficulty in proving the statement for small couplings.
This issue is resolved in \cite{BBL23} by combining trace maps together with a new viewpoint -- applying an interlacing theorem to matrix eigenvalues of the periodic approximations. 
Another crucial ingredient is changing the perspective from considering a single approximation $(\alpha_k)_{k\in\mathbb{N}}$ to the analysis over the whole space of all finite continued fraction expansions simultaneously using a two-level induction.
\end{remark}

\medskip

\subsection*{Step III: The spectrum as the boundary of an infinite tree}
Theorem~\ref{thm-TypeA+B} allows us to construct a (directed) tree $T$ whose vertices represent the spectral bands. The rules that uniquely define $T$ in terms of %a given infinite continued fraction expansion 
$[0,a_1,a_2,a_3,\ldots]$ are sketched in Figure~\ref{Fig-TreeData}.

\medskip

\begin{figure}[htb]
\begin{tikzpicture}[>=stealth,scale=0.75, every node/.style={scale=0.75}]

\begin{scope}[shift={(0,-1)}]
\node at (-5.25,10) {\footnotesize $\sigma(H_{\alpha,V})$};

\foreach \i in {0,1,...,150}
{
\filldraw[black] (9.1*rnd-3.4,10) circle (0.2pt);
}

\foreach \i in {0,1,...,10}
{
\filldraw[black] (0.4*rnd+2.35,10) circle (0.2pt);
}

%%%%%%%%%%%%%%%%%%%%%%%%%%%%%%%%%%%%%%%%%%%%%%%%
\begin{scope}[shift={(0,0.7)}]
\filldraw (-2,6.9) circle (0.5pt);
\filldraw (-2,6.7) circle (0.5pt);
\filldraw (-2,6.5) circle (0.5pt);

\filldraw (0,6.9) circle (0.5pt);
\filldraw (0,6.7) circle (0.5pt);
\filldraw (0,6.5) circle (0.5pt);

\filldraw (2,6.9) circle (0.5pt);
\filldraw (2,6.7) circle (0.5pt);
\filldraw (2,6.5) circle (0.5pt);

\filldraw (5,6.9) circle (0.5pt);
\filldraw (5,6.7) circle (0.5pt);
\filldraw (5,6.5) circle (0.5pt);
\end{scope}
%%%%%%%%%%%%%%%%%%%%%%%%%%%%%%%%%%%%%%%%%%%%%%%%

\end{scope}

\node at (-5.4,5) {\footnotesize $[0,1,2,4]$};
%\filldraw (-3.2,5) circle (1pt);
\node[blue] at (-2.9,5) {\footnotesize $A$};
%\filldraw (-2.6,5) circle (1pt);
\node[blue] at (-2.3,5) {\footnotesize $A$};
%\filldraw (-2,5) circle (1pt);
\node[blue] at (-1.7,5) {\footnotesize $A$};
%\filldraw (-1.4,5) circle (1pt);
\node[blue] at (-1.1,5) {\footnotesize $A$};
%\filldraw (-0.8,5) circle (1pt);
%%%
\node[red] at (0,5) {\footnotesize $B$};
%
%\filldraw (0.8,5) circle (1pt);
\node[blue] at (1.1,5) {\footnotesize $A$};
%\filldraw (1.4,6) circle (1pt);
\node[blue] at (1.7,5) {\footnotesize $A$};
%\filldraw (2.0,5) circle (1pt);
\node[blue] at (2.3,5) {\footnotesize $A$};
%\filldraw (2.6,5) circle (1pt);
%
\node[red] at (3,5) {\footnotesize $B$};
%
%\filldraw (3.4,5) circle (1pt);
\node[blue] at (3.7,5) {\footnotesize $A$};
%\filldraw (4,5) circle (1pt);
\node[blue] at (4.3,5) {\footnotesize $A$};
%\filldraw (4.6,5) circle (1pt);
\node[blue] at (4.9,5) {\footnotesize $A$};
%\filldraw (5.2,5) circle (1pt);
%
\node[red] at (5.6,5) {\footnotesize $B$};

%%%%%
\draw (-2,4)--(-3.2,5);
\draw[->] (-2,4)--(-2.8,4.8);
\draw (-2,4)--(-2.6,5);
\draw[->] (-2,4)--(-2.3,4.8);
\draw (-2,4)--(-2,5);
\draw[->] (-2,4)--(-1.7,4.8);
\draw (-2,4)--(-1.4,5);
\draw[->] (-2,4)--(-1.2,4.8);
\draw (-2,4)--(-0.8,5);
%%%
\draw[->] (1,3.8)--(0,4.8);
\draw (2,4)--(0.8,5);
\draw[->] (2,4)--(1.2,4.8);
\draw (2,4)--(1.4,5);
\draw[->] (2,4)--(1.7,4.8);
\draw (2,4)--(2,5);
\draw[->] (2,4)--(2.3,4.8);
\draw (2,4)--(2.6,5);
\draw[->] (3,3.8)--(3,4.8);
\draw (4,4)--(3.4,5);
\draw[->] (4,4)--(3.7,4.8);
\draw (4,4)--(4,5);
\draw[->] (4,4)--(4.3,4.8);
\draw (4,4)--(4.6,5);
\draw[->] (4,4)--(4.8,4.8);
\draw (4,4)--(5.2,5);
\draw[->] (5,3.8)--(5.6,4.8);
%%%%%

\node at (-5.3,3.8) {\footnotesize $[0,1,2]$};
\node[red] at (-2,3.8) {\footnotesize $B$};
%
%\filldraw (1,3.8) circle (1pt);
\node[blue] at (2,3.8) {\footnotesize $A$};
%\filldraw (3,3.8) circle (1pt);
\node[blue] at (4,3.8) {\footnotesize $A$};
%\filldraw (5,3.8) circle (1pt);

%%%%%
\draw[->] (-2,2.6)--(-2,3.6);
\draw (3,2.6)--(1,3.8);
\draw[->] (3,2.6)--(2,3.6);
\draw (3,2.6)--(3,3.8);
\draw[->] (3,2.6)--(4,3.6);
\draw (3,2.6)--(5,3.8);
%%%%%

\node at (-5.2,2.4) {\footnotesize $[0,1]$};
%\filldraw (-2,2.4) circle (1pt);
\node[red] at (3,2.4) {\footnotesize $B$};

%%%%%
\draw (-2,1.2)--(-2,2.6);
\draw[->] (3,1)--(3,2.2);
%%%%%

\node at (-5.1,1) {\footnotesize $[0]$};
\node[blue] at (-2,1) {\footnotesize $A$};
%\filldraw (3,1) circle (1pt);

%%%%%
\draw (1,0.0) -- (-2,0.8);
\draw (1,0.0) -- (3,1);
%%%%%

\draw[very thick, brown] (1,0) -- (3,1)--(3,2.1);
\draw[very thick, brown] (3,2.6)--(2.1,3.5);
\draw[very thick, brown] (2,4)--(2.25,4.7);

\begin{scope}[shift={(0,-1)}]
\draw[very thick, brown] (2.3,6.165) -- (2.25,6.7) -- (2.4,7.3) -- (2.55,8)--(2.43,8.5)--(2.47,9.1)--(2.45,9.5)--(2.46,9.8)--(2.455,10);
\end{scope}

\node at (-5,0) {\footnotesize root};
\filldraw (1,0) circle (1pt);

\end{tikzpicture}
\caption{An example of a tree for the continued fraction expansion $[0,1,2,4,\ldots]$ is sketched. 
A specific infinite path starting at the root and approaching a point in the spectrum is indicated in brown (see Proposition~\ref{prop-Encoding}).}
\label{Fig-TreeEncoding}
\end{figure}
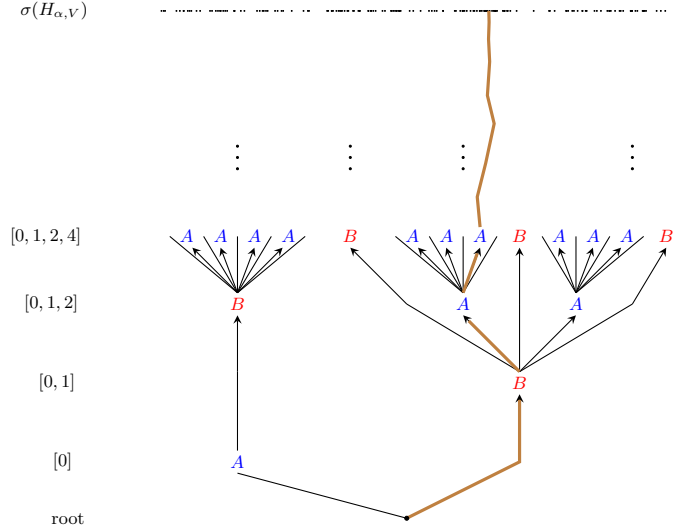

Specifically, each vertex has a type (A or B) according to the spectral band it represents, which is well-defined by Theorem~\ref{thm-TypeA+B}~(a). 
The root of this tree and the two vertices to which the root is connected are sketched in Figure~\ref{Fig-TreeData}~(a).
Then, for each vertex at level $k$ (which is a spectral band $I(V)$ of $H_{\alpha_k,V})$), we connect it with a directed edge to each of the vertices (spectral bands) $J^{(1)},\ldots, J^{(p)}$ of type A and $K^{(1)},\ldots, K^{(p+1)}$ of type B, see Theorem~\ref{thm-TypeA+B}.
We emphasize that the combinatorial data in Theorem~\ref{thm-TypeA+B} -- leading to the construction of the tree -- depends only on the infinite continued fraction expansion of $\alpha$ and is independent of $V>0$. An example of such a tree is provided in Figure~\ref{Fig-TreeEncoding}.

\medskip

We denote by  $\gamma:=(I_{-1},I_0,I_1,I_2,I_3,\ldots)$ an infinite path that starts at the root $I_{-1}=\R$, where $I_m$ are vertices in $T$ satisfying $I_m\to I_{m+1}$ (meaning $I_{m+1}$ is strictly included in $I_m$) for all $m\geq -1$.
Then the boundary of the tree $\partial T$ is the set of all such infinite paths. We have the following correspondence between $\partial T$ and $\sigma(H_{\alpha,V})$.

\begin{proposition}
\label{prop-Encoding}
Let $V> 0$ and $\alpha\in[0,1]\setminus\Q$. Then there exists a surjective map
$$
E_V:\partial T \to \sigma(H_{\alpha,V}), \quad \gamma \mapsto E_V(\gamma).
$$
%is well-defined and surjective.
\end{proposition}

\medskip

\noindent {\bf Proof outline.} 
We start by making the following observation. Even though the combinatorial structure of the tree does not depend on $V>0$, each vertex $I$ in its own represents a particular spectral band $I(V)$ depending on $V$. We now explicitly construct the map $E_V$.

\medskip

Let $V> 0$ and $\gamma=(I_{-1},I_0,I_1,I_2,\ldots)\in\partial T$. Since $\gamma$ is a path, we have $I_m\to I_{m+1}$ for all $m\geq -1$. By definition of the tree $T$, this implies $I_{m+1}(V)\subseteq I_{m}(V)$ for all $m\geq -1$. Thus, the intersection $\bigcap_{m\geq -1} I_m(V)$ is non-empty. By Proposition~\ref{prop-SpectralApprox}, this intersection is contained in $\sigma(H_{\alpha,V})$. Actually, this intersection contains exactly one point: assume by contradiction that there are two different points $E_1\neq E_2$ in $\bigcap_{m\geq -1} I_m(V)$. Since we have an intersection of intervals, we conclude 
$$
\text{Leb}\big( \sigma(H_{\alpha,V}) \big) \geq |E_1-E_2|>0,
$$
where $\text{Leb}$ denotes the Lebesgue measure. This contradicts $\text{Leb}\big( \sigma(H_{\alpha,V}) \big)=0$ proven in \cite{BIST89}. Thus, there is a unique $E_V(\gamma)\in \sigma(H_{\alpha,V})$ such that $\cap_{m\geq -1} I_m(V) = \{E_V(\gamma)\}$. This defines our map $E_V$. 

\medskip

Finally, one can conclude the surjectivity of $E_V:\partial T \to \sigma(H_{\alpha,V})$ by the monotone convergence of the spectra, see Proposition~\ref{prop-SpectralApprox}.
\hfill$\Box$

\medskip

We note that classifying spectral bands into a few types and using a tree encoding to identify open gaps was also used recently in \cite[Theorem~1.11]{LiQuXi14} for the period doubling Hamiltonian.

\medskip

\subsection*{Step IV: The IDS as a function of the tree boundary $\partial T$}

We start by observing that in the definition of the IDS,
$$
N_{\alpha,V}(E) := \lim_{n\to\infty} \frac{\sharp \big\{ \lambda \in \sigma\big(H_{\alpha,V}|_{[1,n]}\big) \,\big|\, \lambda\leq E\big\}}{n}, \qquad E\in\R,
$$
the Dirichlet boundary conditions in $H_{\alpha,V}|_{[1,n]}$ can be replaced by periodic boundary conditions.
This change might modify the numerator, but will not affect the limit.
Thus, counting eigenvalues below $E$ may be replaced with counting spectral bands of $H_{\alpha_k,V}$ that lie to the left of $E$. 
Together with some combinatorial information proven in \cite{Raym}, one concludes the following.

\begin{proposition}
\label{prop-IDS}
Let $V> 0$, $\alpha\in[0,1]\setminus\Q$ and $\alpha_k= [0,a_1,\ldots,a_k]=\frac{p_k}{q_k}$. For each $\gamma\in\partial T$, there are $\pi_k(\gamma)\in \{0,\ldots, a_{k+1}\}$ for $k\geq -1,$ such that
$$
N_{\alpha,V}\big(E_V(\gamma)\big) = -\alpha + \sum_{k=-1}^\infty (-1)^k \pi_k(\gamma) (q_k\alpha - p_k).
$$
\end{proposition}

By the right hand side of the previous equation, it is apparent that the value $N_{\alpha,V}\big(E_V(\gamma)\big)$ depends only on $\gamma$ and is independent of $V>0$.
Interestingly, the values $\pi_k(\gamma)$ for $k\geq -1$ are explicitly described by local properties of $\gamma$ at level $k$ of the tree, see \cite{Raym,BBBT23}.

\medskip

\subsection*{Step V: Identifying a gap label within the tree}

With Proposition~\ref{prop-IDS} at hand and knowing $\pi_k(\gamma)$ explicitly, one can inductively prove the following statement.

\begin{proposition}
\label{prop-IDS_gapLabel}
Let $V> 0$, $\alpha\in[0,1]\setminus\Q$ and $n\in\Z$. Then there exist two different paths $\gamma,\eta\in \partial T$ such that
$$
N_{\alpha,V}\big(E_V(\gamma)\big) = N_{\alpha,V}\big(E_V(\eta)\big) = \{n\alpha\}.
$$
\end{proposition}

Recall that the IDS is monotonically increasing, continuous and it is only constant within spectral gaps, confer (IDS1) and (IDS2). Thus, if there are two different $E_1,E_2\in\sigma(H_{\alpha,V})$ satisfying
$$
N_{\alpha,V}\big( E_1 \big) = N_{\alpha,V}\big( E_2 \big) = \{n\alpha\},
$$
then $(E_1,E_2)$ is an open spectral gap of $\sigma(H_{\alpha,V})$ with gap label $\{n\alpha\}$. 
In light of this, Theorem~\ref{thm-DTMP} follows from Proposition~\ref{prop-IDS_gapLabel}, once we prove that
$E_V:\partial T \to \sigma(H_{\alpha,V})$ is injective.

\begin{theorem}
\label{thm-Injectivity}
For all $V> 0$, the map $E_V:\partial T \to \sigma(H_{\alpha,V})$ is injective.
\end{theorem}

For $V>4$, Theorem~\ref{thm-Injectivity} was proven by Raymond \cite{Raym}. The proof of the statement for $V>0$ can be found in \cite{BBL23}. The extra difficulty in extending the result from $V>4$ to $V>0$ comes from possible overlaps of spectral bands, which do not belong to the same path (see also remark after Theorem~\ref{thm-TypeA+B}). Nevertheless, it is possible to control these overlaps uniformly over the whole space of all finite continued fraction expansions.

\medskip

\medskip

\noindent {\bf Proof of Theorem~\ref{thm-DTMP}.} 
As was already mentioned above Theorem~\ref{thm-DTMP} follows for $V>0$ by combining Proposition~\ref{prop-IDS_gapLabel} and Theorem~\ref{thm-Injectivity}. In order to extend this result to $V<0$, one observes that the spectrum $H_{\alpha_k,V}$ is antisymmetric with respect to $V\mapsto -V$. This yields now
$
\sigma(H_{\alpha,-V})= - \sigma(H_{\alpha,V})
$
and 
$
N_{\alpha,-V}(-E) = 1- N_{\alpha,V}(E)
$, which completes the proof of Theorem~\ref{thm-DTMP}.
\hfill$\Box$

%%%%%%%%%%%%%%%%%%%%%%%%%%%%%%%%%%%%%%%%%%%%%%%%%%%%%
%%%%%%%%%%           References            %%%%%%%%%%
%%%%%%%%%%%%%%%%%%%%%%%%%%%%%%%%%%%%%%%%%%%%%%%%%%%%%
\bibliographystyle{amsalpha}
\bibliography{references}

\providecommand{\bysame}{\leavevmode\hbox to3em{\hrulefill}\thinspace}
\providecommand{\MR}{\relax\ifhmode\unskip\space\fi MR }
% \MRhref is called by the amsart/book/proc definition of \MR.
\providecommand{\MRhref}[2]{%
  \href{http://www.ams.org/mathscinet-getitem?mr=#1}{#2}
}
\providecommand{\href}[2]{#2}
\begin{thebibliography}{BBBT23}

\bibitem[AJ09]{AviJit05}
A.~Avila and S.~Jitomirskaya, \emph{The {T}en {M}artini {P}roblem}, Ann. of
  Math. (2) \textbf{170} (2009), no.~1, 303--342. \MR{2521117}

\bibitem[AYZ23]{AviYouZho23}
A.~Avila, J.~You, and Q.~Zhou, \emph{{D}ry {T}en {M}artini {P}roblem in the
  non-critical case}, 2023, arXiv:2306.16254.

\bibitem[BBBT23]{BBBT23}
R.~Band, S.~Beckus, B.~Biber, and Y.~Thomas, \emph{Sturmian {H}amiltonians for
  large couplings - a review of a work by {L}.~{R}aymond}, 2023+, In
  preparation.

\bibitem[BBG92]{BeBoGh91-Gap}
J.~Bellissard, A.~Bovier, and J.-M. Ghez, \emph{Gap labelling theorems for
  one-dimensional discrete {S}chr\"odinger operators}, Rev. Math. Phys.
  \textbf{4} (1992), no.~1, 1--37. \MR{1160136}

\bibitem[BBL23]{BBL23}
R.~Band, S.~Beckus, and R.~Loewy, \emph{{T}he {D}ry {T}en {M}artini {P}roblem
  for {S}turmian {H}amiltonians}, 2023+, In preparation.

\bibitem[BCDS]{MFO_Bec23}
M.~Baake, M.~I. Cortez, D.~Damanik, and N.~Strungaru, Proceeding of the MFO
  Workshop 2335: Aspects of Aperiodic Order, 27~Aug. - 1 Sep. 2023.

\bibitem[BIST89]{BIST89}
J.~Bellissard, B.~Iochum, E.~Scoppola, and D.~Testard, \emph{Spectral
  properties of one-dimensional quasi-crystals}, Comm. Math. Phys. \textbf{125}
  (1989), no.~3, 527--543. \MR{1022526 (90m:82043)}

\bibitem[BIT91]{BIT91}
J.~Bellissard, B.~Iochum, and D.~Testard, \emph{Continuity properties of the
  electronic spectrum of {$1$}{D} quasicrystals}, Comm. Math. Phys.
  \textbf{141} (1991), no.~2, 353--380. \MR{1133271 (92j:81037)}

\bibitem[DF22]{DamFil22}
D.~Damanik and J.~Fillman, \emph{One-dimensional ergodic {S}chr\"{o}dinger
  operators---{I}. {G}eneral theory}, Graduate Studies in Mathematics, vol.
  221, American Mathematical Society, Providence, RI, [2022] \copyright 2022.
  \MR{4567742}

\bibitem[DG11]{DaGo11}
D.~Damanik and A.~Gorodetski, \emph{Spectral and quantum dynamical properties
  of the weakly coupled {F}ibonacci {H}amiltonian}, Comm. Math. Phys.
  \textbf{305} (2011), no.~1, 221--277. \MR{2802305 (2012g:81065)}

\bibitem[DGY16]{DaGoYe16}
D.~Damanik, A.~Gorodetski, and W.~Yessen, \emph{The {F}ibonacci {H}amiltonian},
  Invent. Math. \textbf{206} (2016), no.~3, 629--692. \MR{3573970}

\bibitem[LQX22]{LiQuXi14}
Q.~Liu, Y.~Qu, and Y.~Xiao, \emph{The spectrum of period-doubling
  {H}amiltonian}, 2022, pp.~1039--1100. \MR{4470245}

\bibitem[Mei14]{Mei14}
M.~Mei, \emph{Spectra of discrete {S}chr\"{o}dinger operators with primitive
  invertible substitution potentials}, J. Math. Phys. \textbf{55} (2014),
  no.~8, 082701, 22. \MR{3390728}

\bibitem[MWL]{MWL23}
\emph{This result was first announced on 1st of {F}ebruary 2023 together with
  the main steps of the proof in the {M}aria-{W}eber lecture of {S}iegfried
  {B}eckus at the {P}otsdam university},
  https://www.math.uni-potsdam.de/institut/veranstaltungen/details-1/veranstaltungsdetails/dry-ten-martini-problem-for-sturmian-dynamical-systems.

\bibitem[Ray97]{Raym}
L.~Raymond, \emph{A constructive gap labelling for the discrete {S}chr\"odinger
  operator on a quasiperiodic chain}, 1997, Preprint.

\bibitem[Sim82]{Sim82}
B.~Simon, \emph{Almost periodic {S}chr\"{o}dinger operators: a review}, Adv. in
  Appl. Math. \textbf{3} (1982), no.~4, 463--490. \MR{682631}

\end{thebibliography}

\end{document}